\documentclass[sigconf,nonacm]{acmart}
\usepackage{algorithm}
\usepackage{algorithmicx}
\usepackage{algpseudocode}         % Load the algorithmicx package for pseudocode (more flexible than algorithmic)
\usepackage{graphicx} % This package is required for \resizebox
\DeclareUnicodeCharacter{0301}{\'{}} % This attaches the acute accent to the preceding character

\AtBeginDocument{%
  }

\begin{document}

\author{Ethan Decker}
\email{ecd5249@upenn.edu}
\orcid{}
\affiliation{%
  \institution{University of Pennsylvania}
  \city{Philadelphia}
  \state{Pennsylvania}
  \country{USA}
}

\renewcommand{\shortauthors}{Ethan Decker}

 \title{Arctic: A Field Programmable Quantum Array Scheduling Technique}


%% Arcticle.

%% Arcticle.
\begin{abstract}

\textbf{Advancements in neutral atom quantum computers have positioned them as a valuable framework for quantum computing, largely due to their prolonged coherence times and capacity for high-fidelity gate operations. Recently, neutral atom computers have enabled coherent atom shuttling to facilitate long-range connectivity as a high-fidelity alternative to traditional gate-based methods. However, these inherent advantages are accompanied by novel constraints, making it challenging to create optimal movement schedules.}

\textbf{In this study I present, to the best of my knowledge, the first compiler pass designed to optimize reconfigurable coupling in zoned neutral atom architectures, while adhering to the reconfigurability constraints of these systems. I approach qubit mapping and movement scheduling as a max-cut and layered cross-minimization problem while enhancing support for spatially complex algorithms through a novel "stacking" feature that balances the qubit array's spatial dimensions with algorithmic parallelism.}

\textbf{I compare the method across various algorithms sourced from Supermarq and Qasmbench where the compiler pass represents the first exclusively movement-based technique to achieve compilation times consistently within seconds. Results also demonstrate that the approach reduces pulse counts by up to 5x and increases fidelity by up to 7x compared to existing methods on currently available technology. }

\end{abstract}



\maketitle

\section{Introduction}
Quantum computers are being developed using various technologies \cite{gonzalez-zalba_scaling_2021} \cite{google_quantum_ai_suppressing_2023},\cite{png_quantum_2022}, \cite{rajak_quantum_2023} with some more promising than others. In recent years, neutral atoms have seen significant advancements, drawing increased attention due to their long coherence times and rapid progress in scalability \cite{bluvstein_quantum_2022}, \cite{tan_qubit_2022}, \cite{bluvstein_logical_2024}. More specifically, the optical properties of neutral atoms facilitate the challenging yet feasible task of scaling qubit arrays to sizes in the thousands \cite{manetsch_tweezer_2024}. Likewise significant, coherence times can be scaled to the 10s of seconds \cite{barnes_assembly_2022} making this platform inherently promising. 

Exploration of software designed for neutral atoms initiated with Baker's seminal paper highlighting the potential for increased connectivity in fixed arrays \cite{baker_exploiting_2021}. Since then, researchers have been drawn to neutral atoms due to their unique properties, such as native multi-qubit interaction gates \cite{patel_geyser_2022} and the foundational technology of optical trapping, which supports programmable array topologies \cite{patel_graphine_2023}. 

Currently, there is an ongoing paradigm shift in neutral atom computing with hardware increasingly focused on the capacity to reconfigure the neutral atom array during algorithm execution to facilitate long-range interactions \cite{bluvstein_quantum_2022}. This reconfigurability—shifting atoms from one spatial position to another—presents an alternative method to gate based routing methods. Experimental demonstrations of atom movement, both prior to and during algorithm execution, have achieved fidelities $\geq 0.9999$ \cite{bluvstein_quantum_2022}. Consequently, this capability of reconfiguration may enable higher algorithmic fidelity compared to those possible with fixed array architectures due to the removal of expensive gate decomposition.

The potential for reconfigurable arrays to achieve high fidelity and low time complexity is evident; however, this approach also introduces a unique set of constraints. Specifically, atom arrays are constructed with two types of optical traps: stationary or mobile. While stationary traps support any 2D topology, mobile traps are limited to a 2D rectangular array configuration. Furthermore, in mobile traps, atoms are moved as entire rows or columns, rather than individually. A second restriction is that mobile traps cannot cross over each other when moved, which only allows for stretching and compressing the grid \cite{tan_qubit_2022}. Thus, both static and mobile traps, coupled with the capability to transfer atoms between these traps, are essential for achieving universal computation. The novelty of these constraints adds an additional layer of complexity to the problem. 

Due to the unique constraints of neutral atom movement, it has been difficult for current techniques to fully optimize compilation centered on reconfigurability. Two aspects in which current techniques struggle are in detaching from dependence on the swap gate, and time complexity when performing the compilation itself. Furthermore, the seminal work of this field \cite{tan_qubit_2022} takes on the order of days to complete due to the need for a solution to NP-hard problems in the execution flow. Other works have since improved on the time complexity \cite{wang_fpqa-c_2023},\cite{wang_q-pilot_2023} however, they do so with the tradeoff of requiring swap-based connectivity and routing algorithms for arbitrary compilation. Importantly, both solutions have great fidelity-enhancing results when comparing to fixed array architectures even with large tradeoffs made. 

\begin{figure}
    \centering
    \includegraphics[width=1\linewidth]{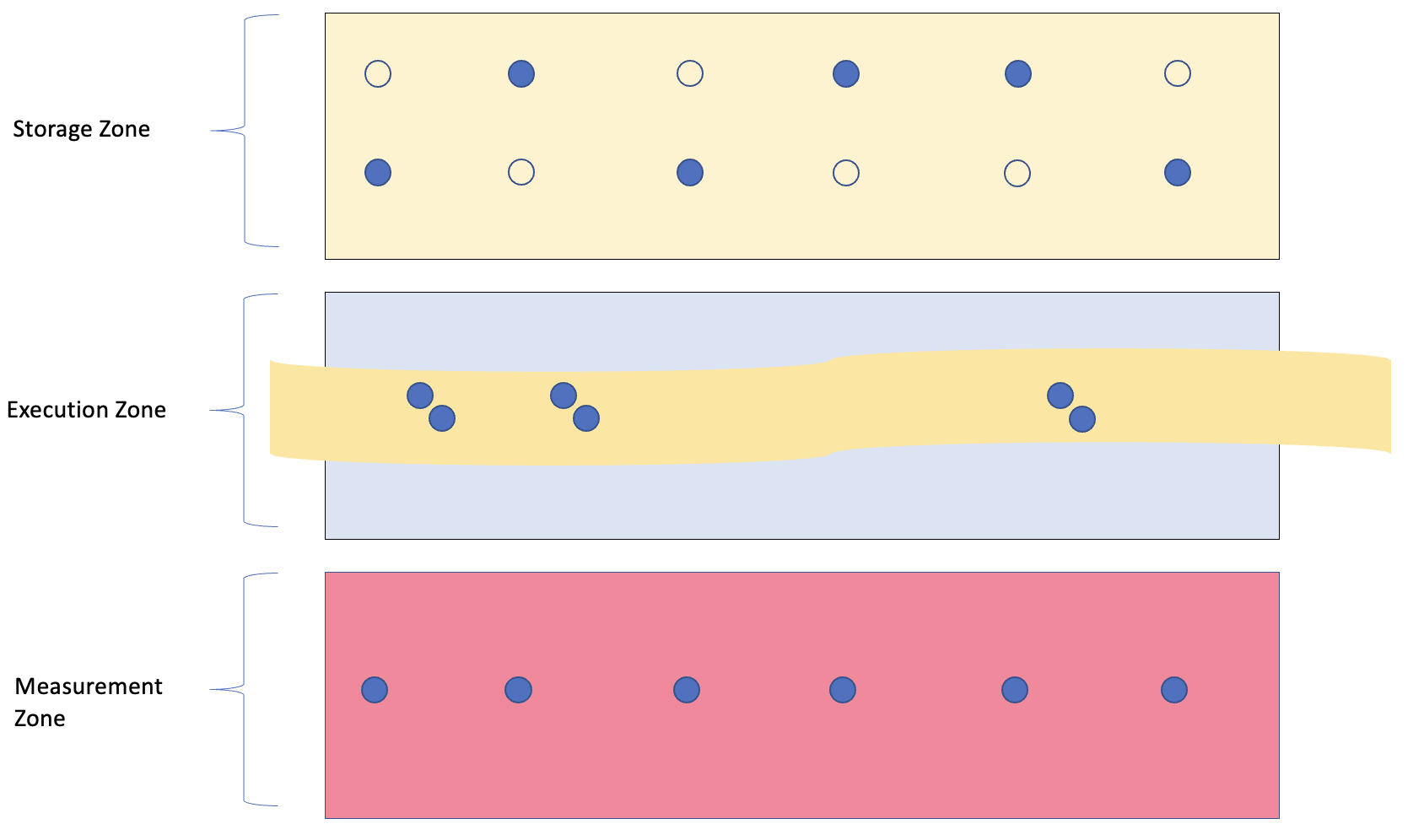}
    \includegraphics[width=1\linewidth]{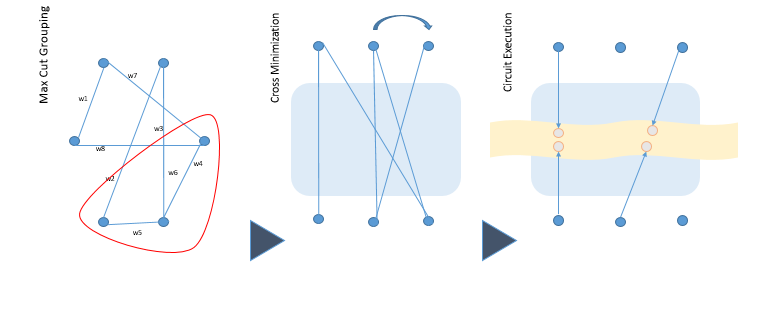}
    \caption{ Top) A zoning scheme for an array of neutral atom qubits. Qubits are stored in SLM traps at the start of computation and shuttled from storage to execution for entanglement. One qubit gates can be executed in the storage zone, while mid-circuit measurements can be preformed in the measurement zone. Botttom) The three phases of Arctic: Phase 1) An example of a circuit connectivity graph being partitioned to maximize qubit trajectories from one layer to another (along one dimension). Phase 2) A swapping of horizontal positions to minimize crossing edges that connect the top and bottom layers that were determined in the previous phase. Phase 3) Atoms being shuffled into the execution zone by an AOD to be coupled by the global Rydberg Laser.}
    \label{fig:enter-label}
\end{figure}

This paper's contribution is Arctic; a compilation technique that allows for both departure from the swap gate and compilation times on the order of seconds through a three-phase process. First, Arctic maximizes qubit trajectories to be along one dimension, reducing the problem to optimization of qubit horizontal ordering in a bi-layer graph. Subsequent cross-reduction heuristics then resolve crossings among qubits moving along the same axis. To enhance compatibility with existing experimental setups, I introduce a dimensional parameter that enables a novel trade-off between the number of layers in the array and the algorithm's parallelism. 

Arctic is the only technique that depends solely on movement while compiling in optimal time periods. Furthermore, to the best of my knowledge, Arctic is the first compiler technique designed for zoned architectures. Likewise significant, the results suggest that it can reduce pulse counts by up to 5x and improve fidelity by up to 7x compared to competing movement-based techniques.

\section{Background to Field Programmable Quantum Array Architecture}

Neutral atom arrays have unique capabilities from other quantum platforms, contributing to the intricacies of their compilation. A through description of the following hardware capabilities can be found in these sources:\cite{graham_multi-scale_2023} \cite{wintersperger_neutral_2023} \cite{henriet_quantum_2020}.  First, each atom needs an optical trap when making a qubit array. Furthermore, traps can be mobile or static with arbitrary topologies generated by overlaying a mobile lattice on a static one, controlled by an Acousto-Optic Deflector (AOD) and a Spatial Light Modulator (SLM), respectively. Static traps can be programmably arranged at the initialization of a quantum computer but remain fixed during execution. In contrast, mobile traps, limited to a rectangular layout, can be dynamically repositioned throughout the execution of quantum algorithms. It is important to remember that mobile traps can only move as entire rows or columns, and such rows or columns cannot intersect each other.

In similar fashion, quantum gates have a unique set of characteristics when being operated. Single-qubit gates are targeted individually within the qubit array using a Raman laser, while two-qubit gates are addressed globally with a Rydberg laser that illuminates a substantial area of the computer \cite{levine_parallel_2019}. The atom-laser coupling facilitates interactions only between illuminated atoms within a specified distance of each other, known as the Rydberg radius. 

Recent experimental setups have designated distinct spatial zones \cite{bluvstein_logical_2024} within the neutral atom plane for different functionalities: a storage zone for idling qubits and locally addressed single qubit gates, an entanglement zone illuminated by the Rydberg laser, and a measurement zone, strategically placed to mitigate decoherence effects (see figure 1). The motivation for zoned architectures originates from the need for error mitigation. By designating separate operation zones, qubits have reduced exposure to lasers and light scattering during operations and measurements thereby enhancing overall fidelity. An additional benefit is that zoned architectures have a familiar abstraction to classical computer architecture which allows for ease of reasoning when considering related design problems. 

The creation of a compiler framework for globally addressed systems present a significant challenge due to the need for precise entanglement, which involves spatially pairing or isolating atoms relative to the Rydberg radius ($r_b$). Work dedicated to introducing this problem can be found in \cite{schmid_computational_2024}. Furthermore, the Rydberg radius necessitates frequent rearrangement operations across every layer of the algorithm, adhering to the constraints previously described. To enable general quantum computation, atoms must be transferred between static and mobile traps—a process taking around 300$\mu s$ per transfer \cite{bluvstein_logical_2024}. Given these extended transfer times, optimizing parallelism is crucial, yet this is complicated by the constraints on atom movement.

\section{Design of Arctic}

The Arctic design is based on the principle that parallelism can be optimized by finding qubit mappings compatible with AOD capabilities and reducing cross trajectories along a single dimension of the array. This intuition originated from observations of typical translational slides on qubits observed during AOD movement. A  benefit to maximizing qubit trajectories to be along one direction is its ability to support cross-minimization by adding more value to the horizontal ordering. In making horizontal ordering the first-order problem to solve, there is a natural reduction in complexity induced by hardware constraints.

The approach is done in three phases where each phase provides support in reducing the problem to ordering along a horizontal dimension. Arctic starts with a quantum algorithm expressed in the neutral atom basis and generates a connection graph, considering only two-qubit gates that can execute in parallel with others. Gates that require serialization due to circuit dependencies are excluded. Sequentially, the technique organizes the virtual qubits based on a max-cut solution applied to the connection graph(see figure 1). This organization is then translated into a physical mapping on a two layered graph topology, facilitating entanglement predominantly along the direction perpendicular to both layers—thereby realizing my initial conceptual insight. To minimize crossing qubit trajectories, Arctic determines an optimal horizontal ordering within each layer using a cross-minimization heuristic.  During algorithm execution, layers of atoms are positioned around the execution zone, conducting two movement operations per layer to transfer atoms involved in gate operations into and out of the execution zone, thus minimizing unnecessary laser exposure. Therefore, it is in maximizing the querying size of atoms from the storage zone to the execution zone that reduces time complexity of quantum algorithms. 

\subsection{Layer Assignment}

Layer assignment aims to find a qubit mapping that maximizes the number of inter-layer two-qubit gates (see figure 1). Recall, that this approach is driven by two key observations. The first is that the collective movement of AOD arrays benefit most when entanglement is uniformly directional. By aligning gate connections uniformly along the axis of qubit movement(the direction perpendicular to both layers), entanglement operations become more effective, as moving whole rows or columns will bring more, ideally all, qubits closer to the desired connection being made. The second observation is that the complexity of constraints put soft limits on the optimization of movement scheduling. By ensuring that most gates are not perpendicular to the axis of movement, optimization is largely confined to horizontal positioning, which allows researchers to sidestep the complexity added with multiple dimensions when moving qubits. 

The assignment phase starts once the filtered connection graph has been created. The filtering process ensures that only gate connections that move as a group are considered. Construction of the connection graph excludes inherently serialized gates because their spatial connections do not enhance parallelism, regardless of their spatial layout. Furthermore, including these irrelevant gates in the optimization process can be detrimental, as it might result in inherently serialized gates being assigned across different layers, while gates executed simultaneously remain within the same layer. 

The task of mapping qubits to one of two layers bears similarities to the max-cut problem, yet it poses a challenge due to its NP-hard complexity. Arctic employs a semi-definite linear program first introduced in \cite{goemans_improved_1995}, with a .87 approximation guarantee. Just as critical, the heuristic used has a polynomial time complexity allowing the technique to remain practical at a large scale, tending to one of two current limitations for reconfigurable compilation. the approximated grouping will assign qubits to either layer 1 or 2 and in an ideal solution, it will maximize the entanglement flowing from one layer to the next along the vertical direction of the qubit array. For a full description of the process, see Algorithm 1.

\begin{algorithm}[H] % Use the [H]ere placement to fix the position in the text
\caption{Grouping Phase Procedure} % Add a caption to describe the algorithm
\begin{algorithmic}[1] % Originally intended for enabling line numbering
\Procedure{GroupingPhase}{$quantum\_circuit$}
    \State $groups \gets []$
    
    \State $G \gets \text{get\_all\_gate\_operations}(quantum\_circuit)$
    \State $C \gets \text{generate\_multi\_qubit\_dag}(G)$
    
    \State $graph \gets \text{initialize empty adjacency matrix based on } C$ 
    
    \For{$front\_layer \in C$}
        \If{$\text{len}(front\_layer) > 1$}
            \For{$(i,j) \in front\_layer$}
                \State $graph[i][j] \gets graph[i][j] + 1$
            \EndFor
        \EndIf
    \EndFor
    \State $groups \gets \text{semi\_definite\_maxcut}(graph)$
\EndProcedure
\end{algorithmic}
\end{algorithm}

\subsection{Cross Minimization}

In phase two of Arctic, the quality of optimization is now determined by a horizontal ordering which can reduce crossing qubit trajectories. Recall, trap transfers are necessary for shuttling qubits in and out of the execution zone. Likewise significant, trap transfers are the most time complex operation in a neutral atom computer and each qubit query from the storage zone to the execution zone requires these transfers. A minimization of crosses allows larger queries by the AOD and therefore allows for more operations per timestep. 

In the implementation I employ the GRASP heuristic \cite{laguna_grasp_1999}, which comprises two main phases: construction and improvement. This choice was informed by an assessment of various cross minimization heuristics \cite{sugiyama_methods_1981},
\cite{chimani_sdp_2012}, \cite{mcgeoch_exact_2014},\cite{marti_heuristics_2003} which identified GRASP \cite{laguna_grasp_1999} as offering the best compromise between performance on sparser graphs and computational efficiency. Importantly, there is a rich set of alternative heuristics that can be applied, varying in performance depending on the graph's density, which in the context corresponds to differently characterized algorithms. 

Sparse graphs are a focus for optimization due to their potential for enhancement in contrast to fully connected graphs or denser equivalents. For example, if there is a bi-layer graph with all to all connectivity, then the horizontal swapping of two qubit positions will lead to the same connection structure. A future research avenue is to consider cross minimization that analyzes circuits at the time step level which will break up dense graphs into sparser ones. 

\begin{figure}
    \centering
    \includegraphics[width=1\linewidth]{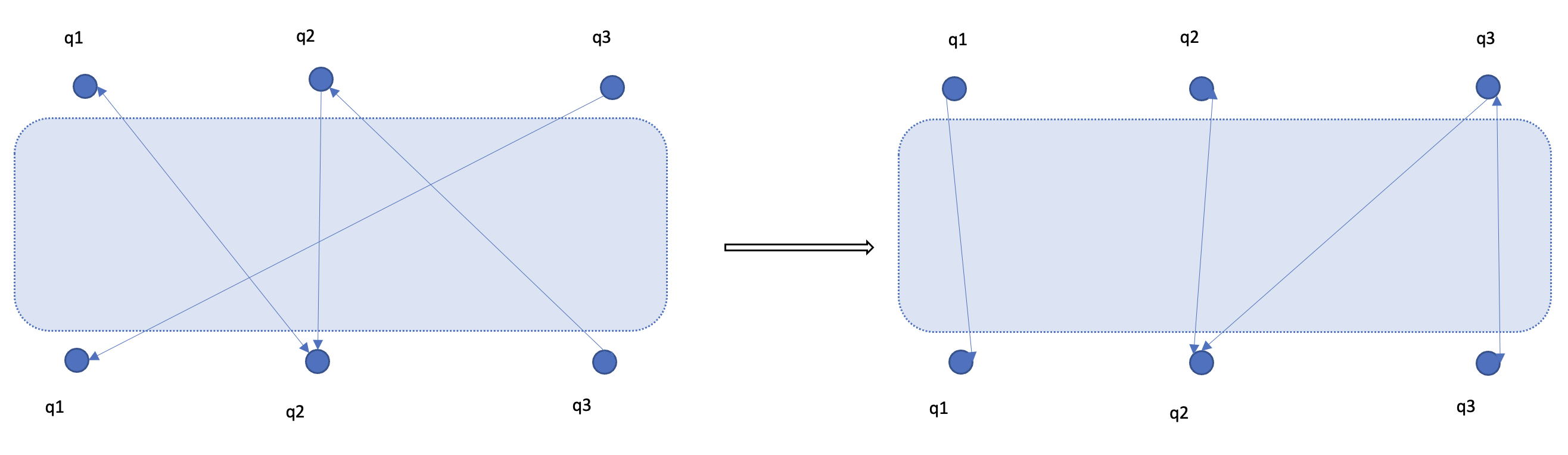}
    \caption{Cross minimization through calculation of the barycenter of neighbors residing in the other layer of qubits. In an ideal scenario, qubits are aligned perfectly with neighbors in the other layer}
    \label{fig:enter-label}
\end{figure}

During the construction phase of the GRASP heuristic, qubits are positioned based on the barycenter of their neighbor's positions. The fundamental premise of this heuristic is that ideally, the barycenter of adjacent qubits in the opposing layer should align the edge representing the two-qubit gate, positioning one qubit in the same column as its partner (see figure 2). Qubits in this phase are analyzed for placement in an order based on their unweighted edge degree, and the barycenter of a qubit not placed is only calculated concerning previously placed qubits. For example, the initial qubit is placed anywhere desired, and the next qubit placement would only be based on whether it is connected to the initial qubit or not. By this ordering, the large graph structure is determined in early iterations and optimization is then left for many of the sparsely connected qubits which would cause damaging serialization.  See Algorithm 2 for a description on the construction phase. 

\begin{algorithm}[H]
\caption{Construction Phase Procedure}
\resizebox{\columnwidth}{!}{ % This command will scale your pseudocode to fit the column width
\begin{minipage}{1.2\columnwidth} % You can adjust the minipage width to slightly more than \columnwidth if necessary
\begin{algorithmic}[1]
\Procedure{Construction Phase}{$qubit\_groups$, $graph$}
    \State $qubit\_positions \gets \text{all qubit horizontal positions}$
    
    \State $Q \gets \text{all placed qubits}$
    \State $U \gets \text{all unplaced qubits}$
    \State $initial\_qubit \gets \text{qubit with highest edge degree}$
    \State $U.\text{remove}(initial\_qubit)$
    \State $Q.\text{append}(initial\_qubit)$
    \State $qubit\_positions\gets \text{initial qubit placement}$
    \While{$\text{len}(U) > 0$}
        \Comment{Select qubit with atleast 2/3 the maximum degree of any qubit at random}
        \State $qubit \gets \text{qubit\_with\_max\_degree\_delta}(graph, Q, \delta= \frac{2}{3}, U)$
        \State $bc \gets \text{barycenter of neighbors in Q}$
        \State $\text{place\_qubit}(qubit, \text{round}(bc))$
        \State $Q.\text{append}(qubit)$
        \State $U.\text{remove}(qubit)$
    \EndWhile
    \State \Return $qubit\_positions$
\EndProcedure
\end{algorithmic}
\end{minipage}
}
\end{algorithm}

In the improvement phase, the initial intuition is refined by making the ordering of qubit placement probabilistic and solely dependent on improvements made. Starting with the qubit embedding produced in the construction phase, Arctic analyzes each qubit to determine if a better placement is possible. This is done by calculating the barycenter of all its now-placed neighbors and examining the cross count when placing the qubit at either the barycenter or the barycenter $\pm 1$. Importantly, a qubit's position is only changed when cross count is reduced. In selecting qubits probabilistically, Arctic employs the distribution below:
$$Pr(q)={P(q,Q) \over \sum_{w \in Q} p(w, Q)}$$
Here, $P(q,Q)$ quantifies the degree of qubit q relative to the group of qubits Q. This formula simulates the practice of choosing qubits with higher degrees first while adding a chance for a sparsely connected qubit to be placed early. Additionally, the probabilistic method of ordering allows for identifying blind spots in placements based on the subgraph and provides the opportunity to adjust the positions of qubits with larger degrees relative to those with smaller degrees. See Algorithm 3 for the implementation of this approach. 

\begin{algorithm}[H]
\caption{Improvement Phase Procedure}
\resizebox{\columnwidth}{!}{ % This command will scale your pseudocode to fit within the column width
\begin{minipage}{1.1\columnwidth} % Slightly larger than \columnwidth to ensure a bit of breathing space
\begin{algorithmic}[1]
\Procedure{ImprovementPhase}{$qubit\_groups ,graph, qubit\_positions$}
    \State $U \gets Q$ \Comment{Prepare for improvement iteration}
    \State $qubit\_probabilties \gets \text{selection\_probabilities(graph, U)}$
    \State $sampled\_indices \gets \text{qubits sorted based on the order sampled}$
    \For{$qubit \text{ in } sampled\_indices$}
        \State $bc \gets \text{barycenter}(vertex, qubit\_positions, Q, graph)$
        \State $\text{improved\_vertex\_placement}(vertex, qubit\_positions, \text{int}(bc))$
    \EndFor
    \State \Return $qubit\_positions$
\EndProcedure
\end{algorithmic}
\end{minipage}
}
\end{algorithm}

GRASP is a heuristic with differing outputs due to the randomness involved and \cite{laguna_grasp_1999} gives a statistical function of iteration outputs for the probability of improvement on the next GRASP iteration. Therefore, it may improve Arctic, but I empirically found 30 to be a good number of iterations to stop at. 

\subsection{Dimensional Variation}
With current neutral atom devices, both dimensions of a rectangular array are limited in size. This is increasingly true as the community moves from computing with physical qubits to logical qubits, where an entire block of atoms is required to represent a single state. To address this, I have introduced a novel "stacking" feature in the methodology, which limits the horizontal dimension of the bi-layer qubit array according to a user-specified parameter. In Arctic's stacking phase, horizontal ordering of qubits is preformed based on the original ordering from the two-layer solution. Vertical ordering of layers is based on the operational frequency of two-qubit gates associated with them. Finally, I propose a new zone topology that has the execution zone in the middle with layers encapsulating it, limiting the distance between the execution zone and the farthest layer.

Introducing more layers to the system results in two primary trade-offs: the emergence of new sources of serialization and the addition of scheduling complexity. Specifically, any edge that lies directly above or below another must be serialized unless it is a direct translation, a situation I have identified for detailed analysis and improvement in future research. Figure 3b illustrates an instance where two-qubit gates, although not intersecting, require serialization. Fortunately, mitigating serialization, due to harsher dimensional constraints,can be done by preserving the initial horizontal order of qubits (Figure 3a). Additionally, by maintaining the ordering it is possible to recycle computational efforts of the cross-minimization phase. Crucially, each new layer will exclusively consist of qubits sourced either from the original top or bottom layer, ensuring the maintenance of valid horizontal ordering throughout the technique's implementation.

The layer assignment for a qubit is dependent on its operational frequency. Sorting qubits in this manner begins by filtering for two-qubit gates that can be executed in parallel with respect to the circuit DAG and the mappings established in the first three phases of the algorithm. Subsequently, qubits are assigned to layers with the layer capacity constrained by the horizontal limit. To maintain high parallelism, the highest operational qubits are placed the closest to the execution zone. For example, if the horizontal dimension limit is 3, then the 6 qubits operated on most frequently, with respect to the filtered gate set, will be in the first two layers (figure 3a). 

\begin{algorithm}[H]
\caption{Stacking Procedure}
\resizebox{\columnwidth}{!}{
\begin{minipage}{1.1\columnwidth}
\begin{algorithmic}[1]
\Procedure{Stack\_Layers}{$qubit\_positions, horizontal\_constraint$, $original\_ordering$}
    \State $layers \gets []$
    \State $gates \gets \text{filter gates based on circuit schedule}$
    \State $top\_qubits \gets \text{layer 1 qubits frequency sorted}$
    \State $bottom\_qubits \gets \text{layer 2 qubits frequency sorted}$

    \State $intermediate\_layer\_top \gets []$
    \State $intermediate\_layer\_bottom \gets []$
    \For{\textbf{each} $(top, bottom) \text{in}(top\_qubits, bottom\_qubits)$}
         \State $intermediate\_layer\_top.\text{append}(top)$
         \State $intermediate\_layer\_bottom.\text{append}(bottom)$

         \If{\text{len}(intermediate\_layer\_top) >= horizontal\_constraint}
            \State $layers.\text{append}(intermediate\_layer\_top)$
            \State $layers.\text{append}(intermediate\_layer\_bottom)$
            \State $intermediate\_layer\_top \gets []$
            \State $intermediate\_layer\_bottom \gets []$
         \EndIf
    \EndFor
    \If{$\text{len}(intermediate\_layer\_top) > 0$}
        \State $layers.\text{append}(intermediate\_layer\_top)$
        \State $layers.\text{append}(intermediate\_layer\_bottom)$
    \EndIf

    \State $layers\_sorted \gets \text{horizontal\_ordering}(layers, original\_ordering)$
    \State \Return $layers\_sorted$
\EndProcedure
\end{algorithmic}
\end{minipage}
}
\end{algorithm}
The layers are vertically arranged in an encapsulating fashion about the execution zone, with the most active layers positioned centrally within the array, adjacent to each other. For instance, if the execution zone begins at vertical position zero, the most frequented pair of layers would be placed at vertical positions +1 and -1, respectively. The next most active layers would be assigned to positions +2 and -2, and this pattern would continue until all layers are positioned, as depicted in figure 3a. My rationale is that such an encapsulation scheme will minimize movement, keeping the high-frequency layers in close proximity to the execution zone to enhance operational efficiency. However, keeping storage of qubits on one side of the execution zone, as traditionally done, is possible by vertically ordering layers based on popularity with the most popular layer closest to the execution zone. 

\begin{figure}
    \centering
    \includegraphics[width=1\linewidth]{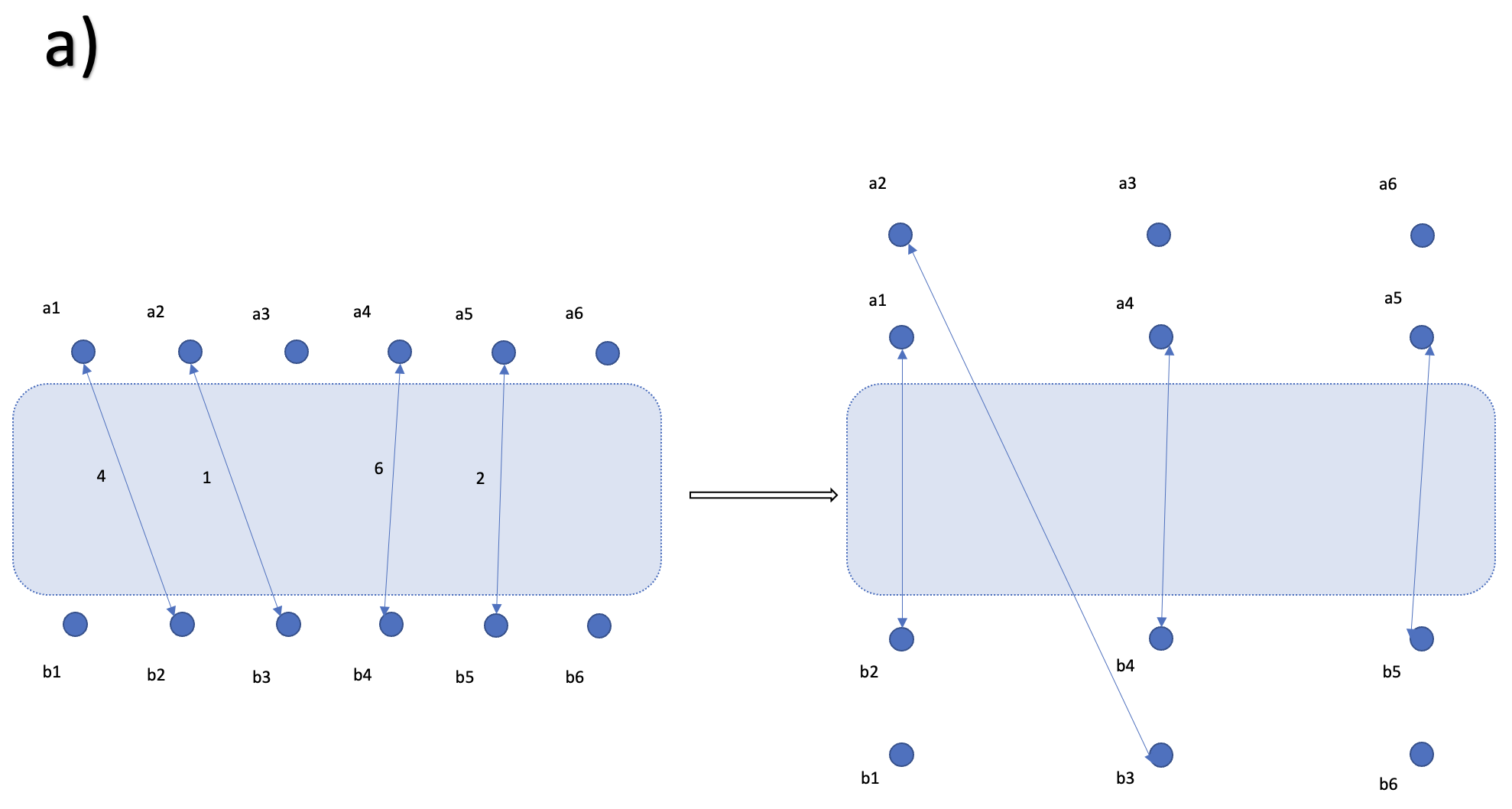}
    \includegraphics[width=1\linewidth]{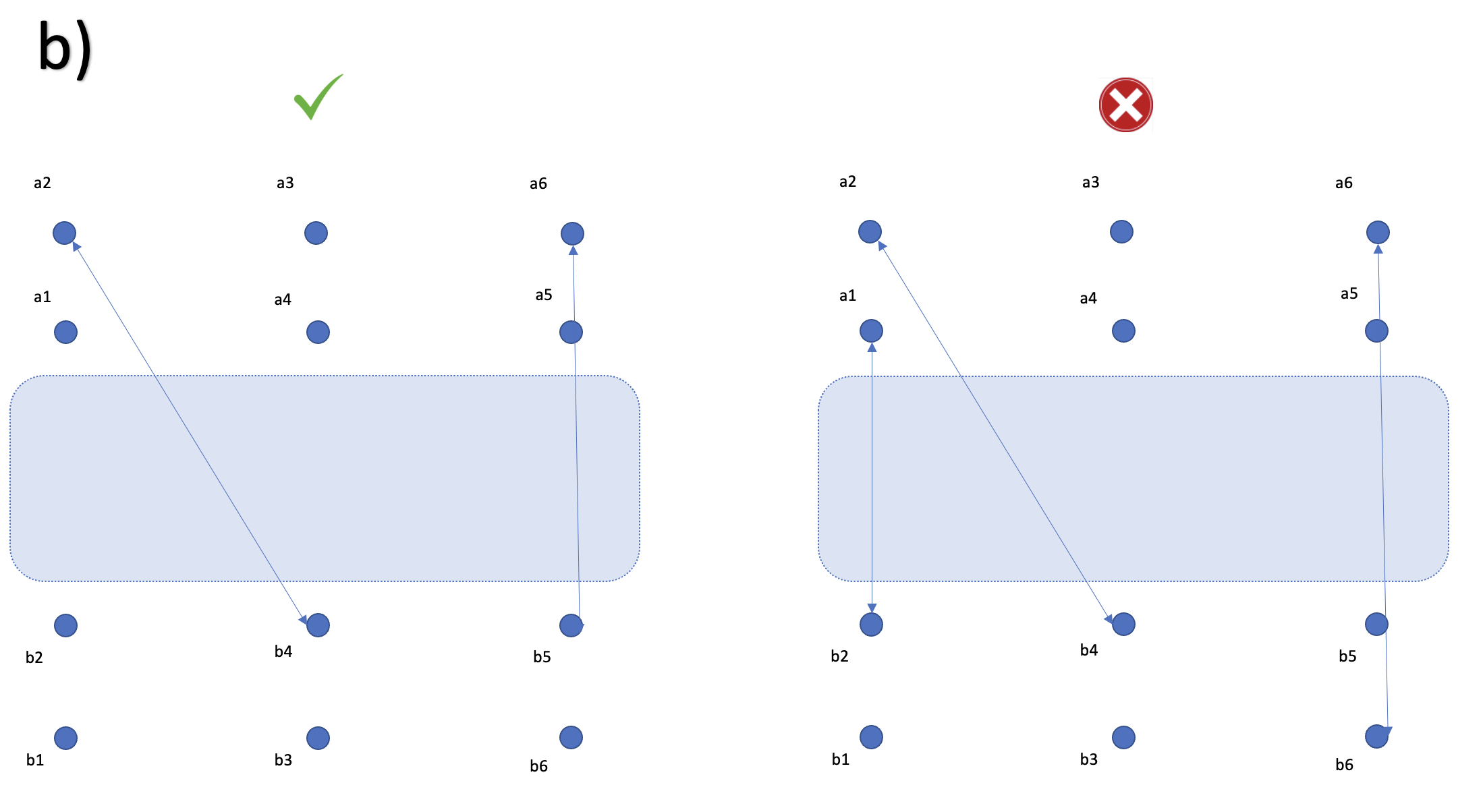}
    \caption{a) A minimized two-dimensional layered graph being transitioned into a four-layer graph where the horizontal ordering between qubits is preserved. Edge weights represent operational frequency, where total edge weighted degree represents the operational frequency of the qubit examined. b) An example of how vertical gates must be serialized due to the constraint that neutral atoms move as whole rows or columns.}
    \label{fig:enter-label}
\end{figure}

\section{Results and Methodology}
To evaluate the technique across both contemporary and futuristic scales, I designed a simulation to assess metrics such as two-qubit gate performance, global pulses, and algorithm fidelity when executed by the technique. For comparative analysis, I benchmarked against FPQA-C \cite{wang_fpqa-c_2023} and a static array of superconducting qubits \cite{arute_quantum_2019}, which represent the state of the art in reconfigurable and fixed arrays. the benchmarking encompassed a broad array of algorithms that varied in width, depth, and connectivity complexity, representative of both current and advanced technological capacities. The evaluation of the technique for any specific algorithm involved recording the best outcomes from 10 runs for each metric. The fidelity model was calculated by treating each source of error as an independent factor and subsequently calculating the product of all error sources. Finally, for the implementation of this simulation, I utilized Python 3.9 and Qiskit 0.44.0 as the software development kit.  

\subsection{Simulating Neutral Atom Architectures}
In the simulation, neutral atom computers are designed to replicate the experimental hardware capabilities at the time of submission. The setup features a single Spatial Light Modulator (SLM) and a single Acousto-Optic Deflector (AOD) dedicated to executing movement and trap transfer operations. 

In this simulation, gate-level errors from both superconducting and neutral atom qubits are treated as independent occurrences. Regarding gate parameters, the fidelity of single-qubit gates for both qubit types is 0.9999 and the fidelity for two-qubit gates is 0.994. Furthermore,the fidelity model for the neutral atom computer is detailed below:
$$F=\Big(\prod_l^L\Big(1- {1-f_2 \over 2}\Big)^{N_l}\Big) \times f_1^{G_1} \times \exp\big(-N {Lt_m \over T_2} \big)$$
This model, adapted from \cite{tan_qubit_2022}, expresses that only laser exposed qubits experience two qubit error. Here, $f_1$ and $f_2$ denote the fidelities of one and two-qubit gates, respectively, with $G_1$ and $L$ representing the number of these gates. The variable $N_l$ indicates the number of qubits exposed at layer $l$ of the algorithm, highlighting that each qubit interacting with the Rydberg laser is susceptible to error.

Decoherence due to qubit movement is described by an exponential term in the equation. The coherence duration is denoted by $T_2=1.5s$ . For temporal complexity, $t_m$ represents the total movement time at each layer, calculated as $t_m = T_0 \sqrt{\frac{D_i}{D_o}}$, where $T_0=300 \mu s$, $D_o=50 \mu s$, and $D_i=30 \mu s$ in the newly conceived zoned architecture \cite{bluvstein_logical_2024} \cite{bluvstein_quantum_2022}. This architecture positions atoms approximately 30$\mu m$ from the execution zone. The variable $N$ represents the total number of qubits in the array, included in the exponential decoherence term. Notably, gate time is excluded from consideration in the neutral atom model due to the significantly greater time involved in trap transfers at each layer, overshadowing error from gate decoherence time.

The computational basis for the neutral atom model, consisting of ['cz', 'u3'], is transpiled using Qiskit version 0.44.0 at optimization level 3 with an all-to-all connectivity backend. 

\subsection{Simulating Superconducting Architecture}
In the simulation of a superconducting model, qubits are connected using a rectangular fixed topology that allows local addressing of any set of gates at each time step during quantum circuit execution. This generous modeling approach is chosen to highlight the negative impacts of gate-based routing on both time complexity and fidelity of futuristic FAA architectures.

For fidelity modeling of superconducting qubits, I utilize a model experimentally validated in studies conducted on the Sycamore quantum processor \cite{arute_quantum_2019} \cite{tan_qubit_2022}. Consistent with the previous approach, I treated each source of error as an independent variable where the fidelity model employed is below:
$$f_1^{G_1} f_2^{G_2} \times \prod_q \Big[1-{1 \over 3} \Big({1 \over T_1} + {1 \over T_2} \Big )T_q^{idle} \Big) \Big]$$

Here, decoherence time scales are represented according to current experimental qubit coherence \cite{arute_quantum_2019}. $T_1=15 \mu s$  and $T_2=25\mu s$ represent the relaxation and dephasing times, respectively, with $T_q$ denoting the total idle time for qubit $q$ during algorithm execution. 

The gate fidelities are annotated as in the previous model, and the execution time per algorithm layer is set at 25$ns$, which corresponds to the duration of the longer gate operation. These parameters are sourced from recent studies modeling experimental setups for superconducting quantum computers \cite{arute_quantum_2019}.

In modeling the Sycamore processor, the basis for superconducting qubits includes [root\_iswap, rx, rz], with the Qiskit transpiler set to level 3 optimization when transpiling and mapping to the square grid connectivity.  

\subsection{Algorithm Benchmark} 

The bedrock motivation for quantum compilation is the aspiration to reliably execute practical quantum algorithms and extract the maximum capability from the hardware. Consequently, I tailored the algorithms based on criteria outlined in Supermarq \cite{tomesh_supermarq_2022} and sourced qasm level expressions of the algorithms from QASMBench \cite{li_qasmbench_2023}. The only exception to the source of algorithms is QAOA where I utilized a custom script to generate circuits of varying depth and size followed by a transpilation step to convert operations to the required computational basis. The types of algorithms in the benchmark include variational, AI-related, arithmetic, and simulation-based, all of which hold near-term potential and challenging algorithm characteristics. 

I adopted various characteristics in the benchmarks that pose challenges in general compilation yet can be enhanced through parallelism. Algorithm attributes diversified include spatial complexity, time complexity and qubit connectivity. Recall that given the variability in both transpilation methods and techniques, the evaluation for fixed and reconfigurable architectures was conducted 10 times. Subsequently, I recorded the optimal results for pulse count, gate count, and fidelity. I also documented compilation times for Arctic to evaluate the practicality at current and prospective scales. Throughout the tests, I observed that compilation times did not exceed two minutes (see Figure 4).

\begin{figure}
    \centering
    \includegraphics[width=1\linewidth]{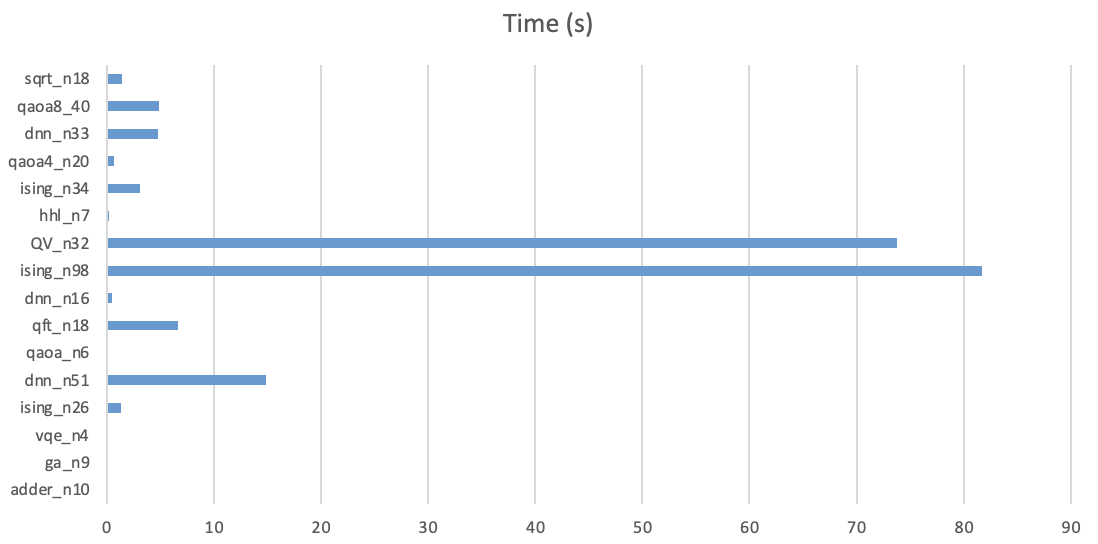}
    \caption{Compilation Time in Seconds}
    \label{fig:enter-label}
\end{figure}

\subsection{Related Work}
\textbf{Fixed Architecture Compilation:} Fixed architectures are quite common, and necessary in superconducting computers\cite{arute_quantum_2019}. In this architecture, qubits are fixed in a spatial location with nearest neighbor connectivity. This paradigm, one of the earliest in quantum computing, offers straightforward constraints of communication however, it faces significant challenges due to the complexity of the routing problem. Extensive research has been devoted to addressing connectivity issues on fixed arrays \cite{li_tackling_2019}, \cite{nottingham_decomposing_2023}, \cite{patel_graphine_2023}, yet algorithms are restricted in fidelity due to swap gate overhead that has damaging effects at larger sizes. Neutral atom devices, with their capability for movement-based connectivity, bypass these limitations obviating the need for complex routing solutions with expensive gate decomposition. 

\textbf{Movement-Based Compilation:} The exploration of movement-based compilation has gained traction recently with \cite{tan_qubit_2022} pioneering this approach through the flexibility of neutral atom arrays. A key innovation of the work is the formulation of reconfigurability constraints into boolean equations which facilitates the integration of familiar computational methods while respecting encoded constraints. This compilation technique offers the advantage of replacing swap gates with less costly movement operations, generally enhancing algorithm fidelity. However, the primary drawback lies in the reliance on SMT-based solvers, an exponentially complex operation, often leading to timeouts and scalability issues. Another issue is the unnecessary exposure of idling qubits on Rydberg lasers, which sacrifices fidelity. In response, the Arctic framework employs a set of heuristics that ensure polynomial execution time while taking advantage of zoning to limit laser exposure. 

\textbf{Hybrid Compilation:} In response to the complexities associated with boolean equations, the field has developed alternative approaches that simplify these issues into more manageable problems \cite{wang_fpqa-c_2023}. Notably, certain studies have identified that algorithms with bipartite connectivity are well-suited to the characteristics of reconfigurable arrays, though most algorithms do not inherently exhibit this structure. By employing max k-cut heuristics, researchers have devised an innovative routing scheme between groups rather than individual quantum registers. After routing is preformed, the work employs a movement scheme that offers scalable compilation times while also leveraging the benefits of movement operations. However, routing between groups requires swap gates and the continued use of swap operations has prevented these methods from fully realizing the potential of movement-only strategies. Additionally, hardware innovations must be made to allow for multiple AOD devices giving way to the full potential max-k cut solutions. Finally, the technique also does not consider unnecessary exposure to a global Rydberg laser which severely limits fidelity saved from movement-based operations. Conversely, Arctic achieves this potential by exclusively utilizing movement for full connectivity and maintains competitive compilation times on current experimental apparatuses with one AOD, while being able to benefit from multiple AOD components as they are incorporated into future devices. Recall, Arctic is also conscientious of exposure to the Rydberg laser by employing advantages of zones on the neutral atom array not seen in techniques before. 

\subsection{Measured Results}

\textbf{Gate Count}: The first metric assessed is the total number of gates required across all three compilation modes. As illustrated in Figure 5, this metric highlights the overhead associated with routing and its gradual reduction through movement-based operations. Algorithms that are deeper and have greater connectivity exhibit an exponential increase in gate counts, underscoring the complexity of routing. In contrast, Arctic maintains a constant gate count post-transpilation into the system basis, significantly reducing error introduction.

\textbf{Pulse Count}: The second metric examined is global pulse count, relevant exclusively to neutral atoms; thus, layers of an FAA are not considered. Figure 6 compares Arctic, both with and without horizontal constraints, to FPQA-C. An increase in pulses is anticipated as dimensional constraints become more stringent due to the additional serialization from vertical constraints. Additionally, FPQA-C tends to be very competitive on very dense connection graphs however, Arctic commonly outperforms FPQA-C by at least 50 percent across most algorithm types and in best cases, by a multiple of 5x. 

\begin{figure}
    \centering
    \includegraphics[width=1\linewidth]{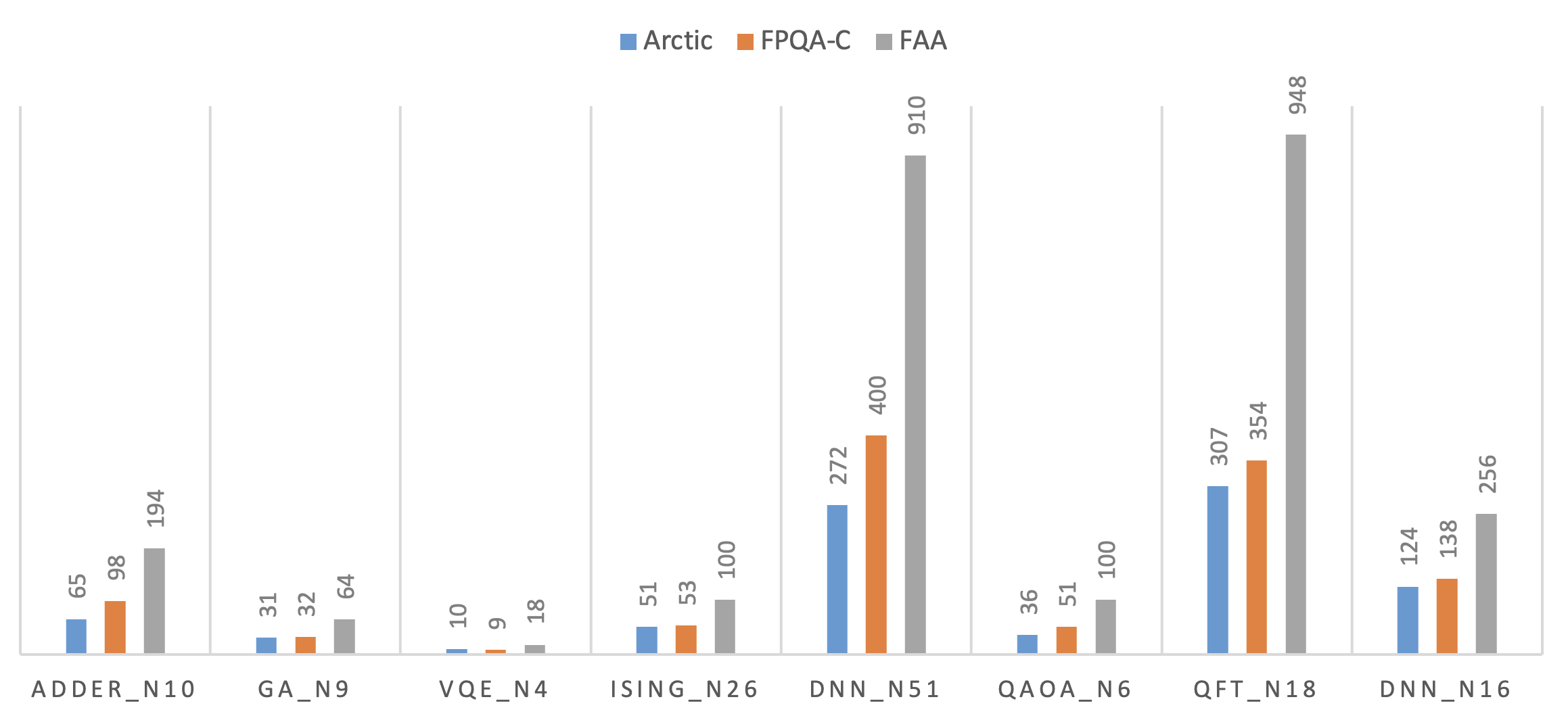}
    \includegraphics[width=1\linewidth]{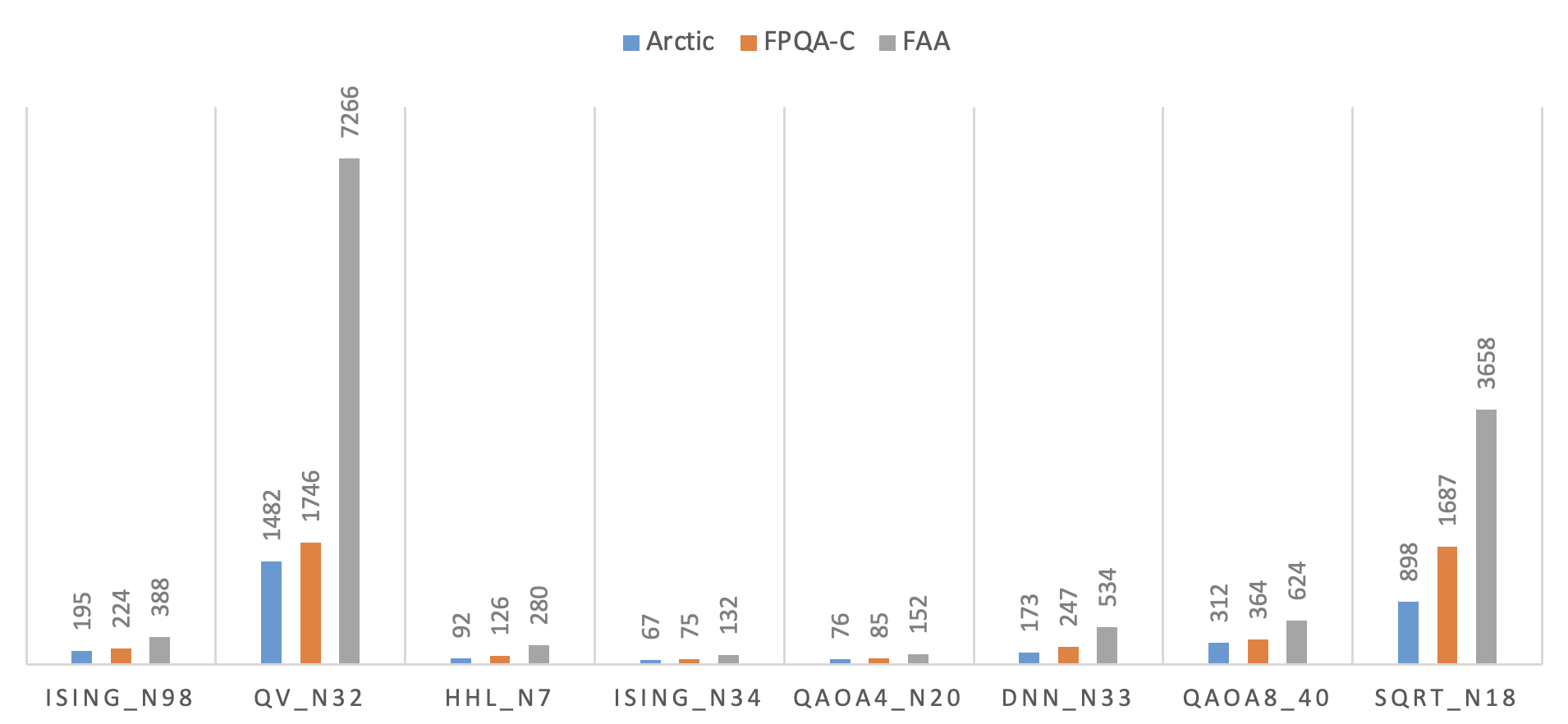}
    \caption{Number of Two Qubit Gates}
    \label{fig:enter-label}
\end{figure}

\begin{figure}
    \centering
    \includegraphics[width=1\linewidth]{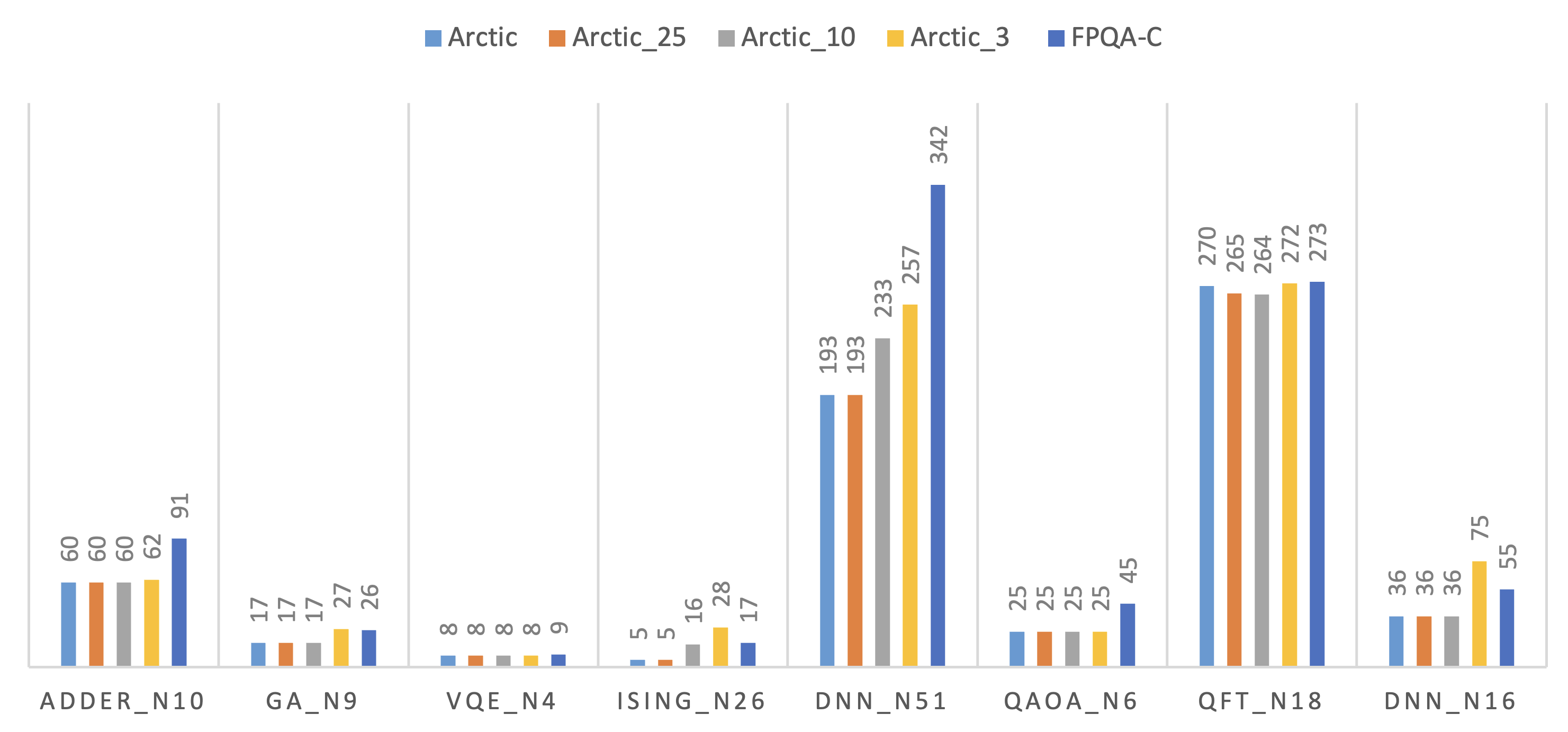}
    \includegraphics[width=1\linewidth]{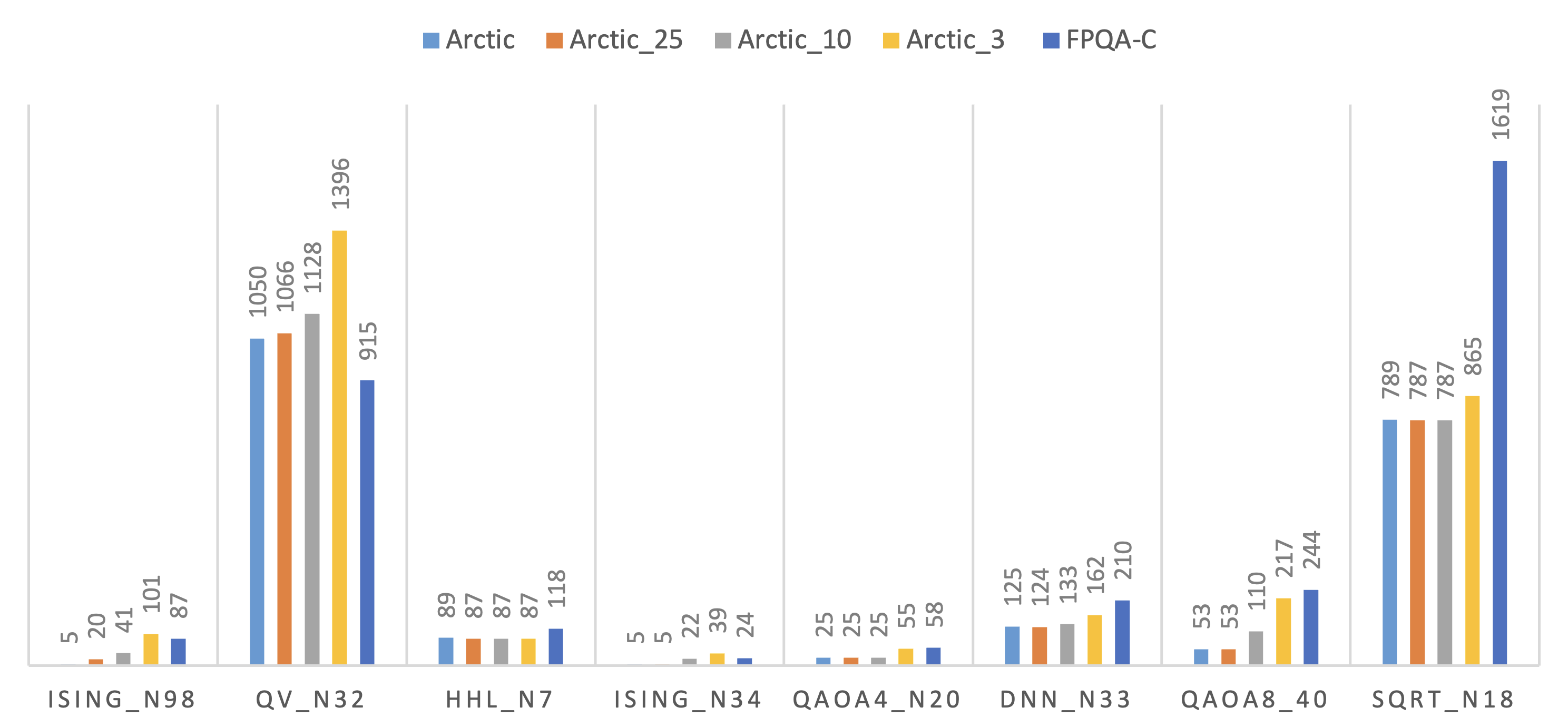}
    \caption{Number of Two Qubit Layers}
    \label{fig:enter-lab}
\end{figure}

\begin{figure}
    \centering
    \includegraphics[width=1\linewidth]{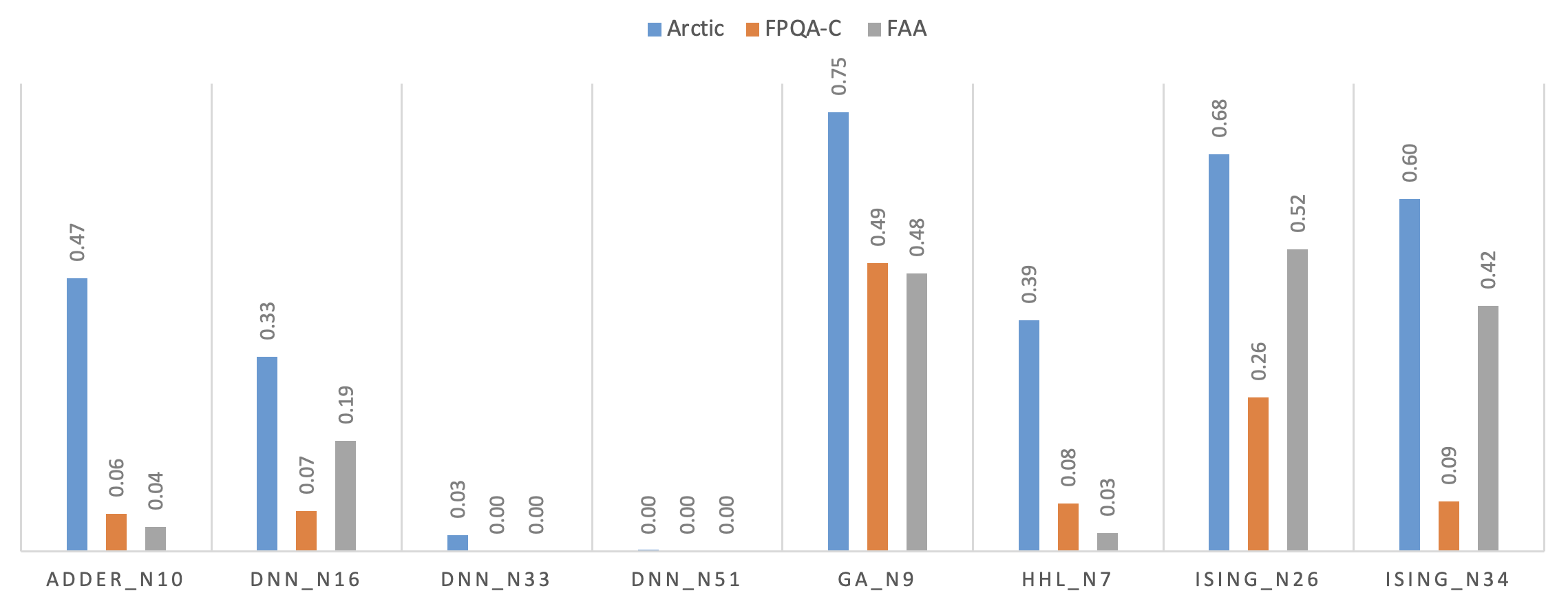}
    \includegraphics[width=1\linewidth]{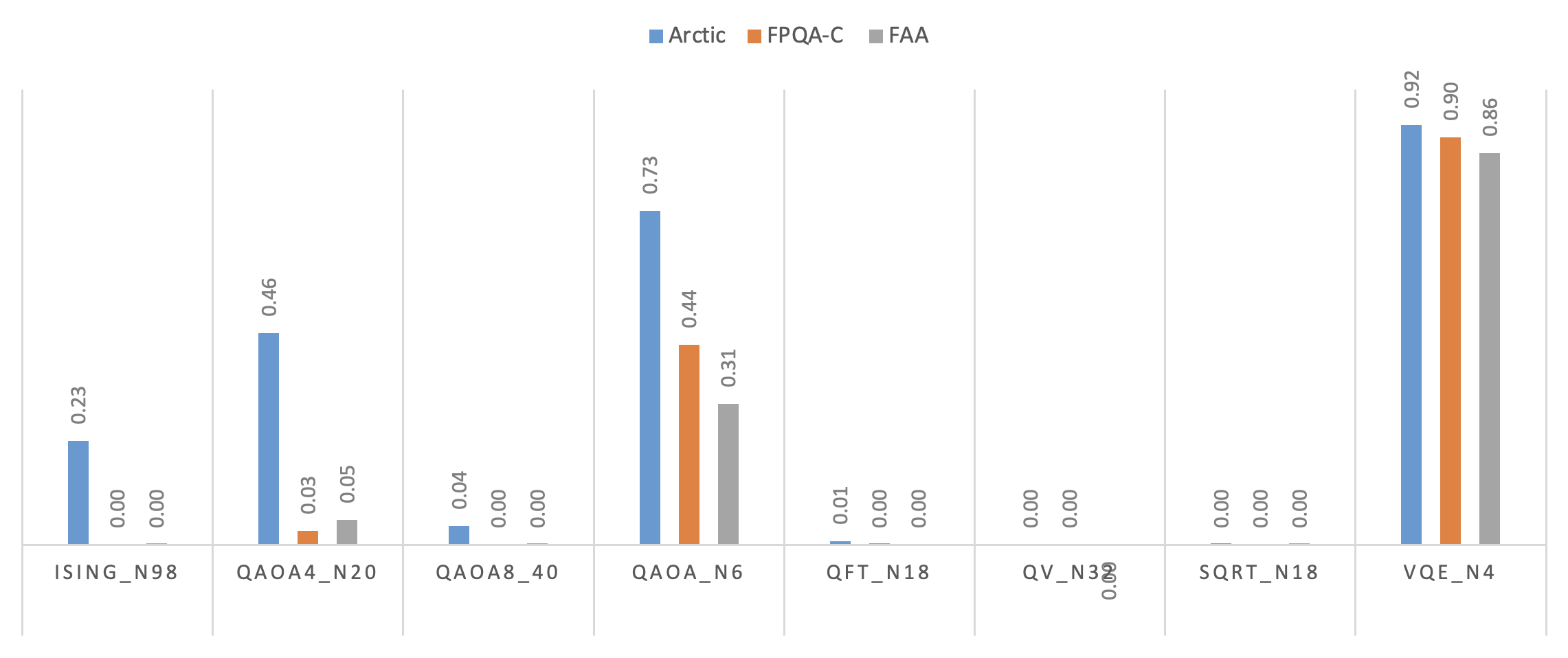}
    \caption{Total Fidelity}
    \label{fig:enter-labe}
\end{figure}

\textbf{Fidelity}: the final evaluation metric is fidelity, depicted in Figure 7. This data demonstrates the detrimental effects of global pulses when atoms are unnecessarily exposed to the Rydberg laser, as observed with FPQA-C. Additionally, the results highlight the damaging effects of routing compared to trap transfer and movement overhead with additional advantages from zoned computers. Conclusively, due to the advantage of limited laser exposure and movement-based routing, the fidelity in the best cases improves the state of the art by a factor of 7x. 

\section{Conclusion}

It is an exhilarating time to be involved in the development of neutral atom computers, which hold immense potential despite their novel constraints. From a software perspective, this geometric compilation challenge represents a groundbreaking problem, offering ample learning opportunities. Beyond academic interest, the elimination of costly swap gates, combined with inherent advantages such as extended coherence times and scalable architectures, underscores the critical need for improved compilation techniques.

Arctic delivers superior outcomes among movement-based compilers however, there is still considerable room for improvement. Crucially, achieving optimal compilation requires harnessing the potential of multi-dimensional configurations. Strategies that extend parallel entanglement beyond a single dimension are essential for the advancement of practical quantum computing. Future innovations might leverage the symmetries of gate edges or employ additional hardware resources, like multiple Acousto-Optical Deflectors to enable movement across both array dimensions. Other interesting avenues for achieving better computation are in regards to multi-specied arrays \cite{singh_dual-element_2022} and three-dimensional grids \cite{barredo_synthetic_2018}.  Given the new zoning abstraction to quantum computers, exploring different zone topologies may lead to better query times or passive error mitigation. 

Quantum Computing stands at a crossroads as a potentially transformative or inconsequential technology. Techniques that take advantage of hardware capabilities are vital for discovering what is possible in computing. Reconfigurable architectures have the potential to be truly great and I hope for continued exploration of this paradigm. 
\section*{Acknowledgements}
I would like to thank Tirthak Patel for the helpful conversations and encouragement to pursue Arctic as a technique. 

\bibliographystyle{acm}
\bibliography{referencesMor}

\end{document}